\documentstyle[prd,epsfig,aps,preprint]{revtex}
\begin{document}

\draft
\title{Power Law Stellar Distributions}
\author{J. A. S. Lima$^{1, 2}$\footnote{limajas@astro.iag.usp.br, limajas@dfte.ufrn.br}
and  R. E. de Souza$^{1, b}$\footnote{ronaldo@astro.iag.usp.br}}

\address{$^{1}$Instituto de Astronomia, Geof\'{\i}sica e
Ci\^encias Atmosf\'ericas, Universidade de S\~ao Paulo\\ Rua do
Mat\~ao, 1226 - Cid. Universit\'aria, 05508-900, S\~ao Paulo, SP,
Brasil\\ $^{2}$Departamento de F\'{\i}sica, Universidade Federal
do Rio Grande do Norte\\ C. P. 1641, 59072-970,  Natal, RN,
Brasil}


  \smallskip

\date{\today}
\maketitle

\vskip 1.5cm

\begin{abstract}The density profiles and other quantities of physical
interest for spherically symmetric systems are computed by
assuming that a collisionless stellar gas may relax to the
non-Gaussian power law distribution suggested by the nonextensive
kinetic theory. There are two different classes of solutions. The
first class behaves like a subset of the polytropic Lane-Emden
spheres, whereas the second one corresponds to a transition
between two different polytropic indices. Unlike the isothermal
Maxwellian sphere, the total mass and sizes of both classes are
finite for a large range of the nonextensive q-parameter.
\end{abstract}


\newpage
\section{Introduction}

It is widely assumed that self-gravitating stellar systems like
globular clusters are completely or almost completely relaxed
because they are changing very slowly, or equivalently, the
characteristic evolution time scale is too long. In particular,
this means that stellar encounters combined with the action of
some long range kinetic processes like ``phase mixing" and
``violent relaxation" have driven their velocity distributions
toward a collisionless kinetic steady state (Spitzer \& H\"arm
1958, King 1962, Michie 1961, 1963, Lynden-Bell 1967, Spitzer
1987). Although considering the numerous and important
contributions to this field, an overlook in the recent literature
suggests that the equilibrium (or near-equilibrium) distribution
of these relaxed systems is still not firmly established (beyond
doubt), and basically remains as an open question (Drukier et al.
1992, Meylan \& Heggie 1997, Katz \& Okamoto 2000, Chavanis 2002).

The phase-space density for spherically symmetric systems was
initially described by the equilibrium Maxwell-Boltzmann (MB)
distribution
\begin{equation}\label{e1}
f({\bf r},v)=\rho_1\left({2\pi\sigma^{2}} \right)^{-3/2}\exp
\left(-{{1\over 2} v^2 + \Phi({\bf r})\over \sigma^{2}}\right)
\end{equation}
where $\Phi$ is the potential per unit mass, $\rho_1$ is the
density in the absence of the force field and $\sigma$ is the
velocity dispersion. However, the computation of the associated
mass density gave rise to a very serious problem, namely: the
total mass is infinite (Chandrasekhar 1960, Binney \& Tremaine
1994). This undesirable feature is more easily seen from the
particular solution, $\rho(r) = {{\sigma}^{2}/2\pi G r^{2}}$
(singular isothermal sphere) which predicts an infinite total
mass. Many ad hoc phenomenological distributions were proposed to
solve this problem with basis on Jeans' theorem: any steady state
solution of the collisionless Boltzmann (or Vlasov) equation
depends on the phase space coordinates only through isolating
integrals of motion (Jeans 1929).

A different route was initiated by Spitzer and H\"arm (1958), and
further worked out by Michie (1961, 1963) and King (1962, 1965,
1966). It was recognized that star clusters cannot be exactly
described by the expected MB equilibrium distribution. However, it
was also assumed that relaxation processes always drive the
stellar distribution as far as it can to a quasi-MB final state
which should be determined, for instance, by solving the
non-equilibrium Fokker-Planck equation. This line of inquiry lead
to some interesting and consistent results among them the lowered
isothermal sphere, as well as the anisotropic models proposed by
Michie.

In this article we consider a different approach.  With basis on
the q-power law equilibrium distribution we determine the radial
and projected density profiles for two large classes of isothermal
stellar systems. It should be recalled that the nonextensive
treatment for a  stellar collisionless systems was first
considered by Plastino \& Plastino (1993) through a variational
principle where the Tsallis (1988) entropy formula was maximized
taking into account the constraints imposed by the total mass and
energy density. Here, we consider directly the nonextensive
distribution which follows naturally from the kinetic equilibrium
q-entropy formula (Lima et al. 2001). An attractive feature of
this power law distribution is that the models are analytically
tractable in such a way that a detailed comparison with the
standard Maxwell-Boltzmann approach is immediate. As we shall see,
there are at least two classes of solutions with finite mass and
radius. Actually, the first class is not new, however we discuss
some related subtleties not considered in the quoted articles.

The paper is structured as follows. Next section we set up the
basic equations and discuss the density profiles to the first
class of nonextensive spherically symmetric self-gravitating
stellar systems based on Tsallis' distribution. Section 3 discuss
the main features of the trunctated models, and, in section 4, we
resume briefly the main results.

\section{Power Law Stellar Spheres}

Let us now consider the following non-Gaussian velocity
distribution

\begin{equation}\label{e2}
f(v) = {{\rho_1 C_{q}}\over {{(2\pi \sigma)}^{3/2}}}\left[1 -
(1-q){v^2\over 2\sigma^{2}}\right]^{1 \over {1-q}}
\end{equation}
where the $q$-parameter quantifies to what extent the distribution
departs from the standard Gaussian form (Silva et al. 1998, Lima
at al. 2001, Mendes \& Tsallis 2001, Kaniadakis 2001). Actually,
it is a power law if $q \neq 1$, whereas for $q=1$ it reduces to
the Maxwellian function. This result follows directly from the
known identity $\rm{lim}_{d \rightarrow 0}(1 + dy)^{1 \over d} =
{\rm{exp}(y)}$ (Abramowitz \& Stegun 1972). The quantity $C_q$ is
a $q$-dependent normalization constant given by (Lima et al. 2002)

\begin{equation}\label{e3}
C_q = (1-q)^{1/2}(\frac{5-3q}{2})(\frac{3-q}{2})\frac{\Gamma
(\frac{1}{2}+{1\over 1-q})}{\Gamma({1\over 1-q})}\,\,\, for \, q
\leq 1
\end{equation}
and
\begin{equation} \label{e4}
C_{q}=(q-1)^{3/2}\left(\frac{3q-1}{2}\right)\frac{\Gamma\left(\frac{1}{1-
q}\right)}{\Gamma\left(\frac{1}{1-q}-\frac{3}{2}\right)} \,\,\ for
\, q \geq 1,
\end{equation}
which reduce to the expected result in the limit $q=1$. For
instance, for $q \leq 1$ we define $z=(1-q)^{-1}$ so that $z
\rightarrow \infty$ if $q$ goes to unity. In addition, from the
identity $\rm{lim}_{\vert z \vert \rightarrow
\infty}z^{-\rm{a}}\Gamma(\rm{a} + z)/\Gamma(z) = 1$ (Abramowitz \&
Stegun 1972), one may see that $C_{1} = 1$, thereby showing that
the normalization of the standard Maxwellian distribution is
recovered in this limit.

The power law distribution (2) can be deduced from two simple
requirements: (i) isotropy of the velocity space, and (ii) a
suitable nonextensive generalization of the Maxwell
factorizability condition, or equivalently, the assumption that in
this enlarged framework $f(v) \neq f(v_x)f(v_y)f(v_z)$. It was
also shown that for $q
> 0$, the above distribution function satisfies a generalized
H-theorem, and its reverse has also been proved, that is, the
collisional nonextensive equilibrium is given by the Tsallis'
power law velocity distribution (Lima et al. 2001). In the last
few years, several applications of this equilibrium power law
velocity distribution have been done in many disparate branches of
physics (Liu et al. 1994, Bhogosian 1996, Lima et al. 2000,
Tsallis et al. 2001). In the astrophysical context, Taruya and
Sakagami (2002), investigated in detail the problem of instability
(gravothermal catastrophe) working in the so-called (u,v) plane
(Milne's variables). Further, they shown that the king model fails
to match the simulated distribution function, especially at
cut-off scales (Taruya \& Sakagami 2003). Even the Jeans
gravitational instability criterion for a collisionless system was
recently discussed in this enlarged framework (Lima et al 2002).
In particular, for power law distribution with cut-off, it was
shown that such a system presents instability even for wavelengths
of the disturbance smaller than the standard critical Jeans value.
For a selfgravitating system, the nonextensive phase-space density
reads (Lima et al. 2002a)

\begin{equation}\label{e5}
f({\bf r},v) = {{\rho_1 C_{q}}\over {{(2\pi
\sigma)}^{3/2}}}\left[1 - (1-q)\left({{1\over 2} v^2 + \phi({\bf
r})\over \sigma^{2}}\right)\right]^{1/(1-q)}
\end{equation}
which can also be obtained by integrating the Vlasov equation. As
should be expected, the above expression reduces to equation (1)
in the limiting case $q=1$.

Let us now consider the velocity distribution (\ref{e5}) to build
a set of stellar systems whose structure is sustained by their
respective gravitational field. Following standard lines, the
granularity of the star system is ignored, the gravitational
potential is assumed to be a slowly varying function of position,
and any change in the physical properties due to collisions or
evolution of the stars are neglected. The radial dependence of the
mass density is obtained by integrating (\ref{e5}) over all
allowed velocities

\begin{equation}\label{e6}
\rho= {4\pi{\rho_1 C_{q}}\over {{(2\pi \sigma)}^{3/2}}} \int
\left[1 - (1-q)({{1\over 2} v^2 + \phi({\bf r})\over
\sigma^{2}})\right]^{1/(1-q)} v^{2}dv
\end{equation}
whereas the gravitational potential $\phi$ must be determined from
Poisson's equation

\begin{equation}\label{e7}
\frac{1}{r^2}\frac{d}{dr}(r^2\frac{d\phi}{dr})= 4\pi G\rho.
\end{equation}

In order to simplify further the integral (\ref{e6}), it is
convenient to introduce the dimensionless energy per unit mass of
a star

\begin{equation}\label{e8}
\varepsilon = {\frac{1}{2}v^2 + \phi \over \sigma^{2}}
\end{equation}
and from the argument of the power law in (\ref{e5}), this
quantity is restricted by $\varepsilon \leq \frac{1}{1-q}$, in
order to guarantee a definite real valued distribution function
(unless explicitly stated, in what follows we consider only the
case $q \leq 1$). Now, inserting $\varepsilon$ into (\ref{e6}) one
has

\begin{equation}\label{e9}
\rho= {4\pi{\rho_1 C_{q}}\over {{(2\pi \sigma)}^{3/2}}}
\int_{\phi\over \sigma^{2}}^\frac{1}{1-q} [1 -
(1-q)\varepsilon]^{\frac{1}{1-q}}(\varepsilon -{\phi\over
\sigma^{2}})^{\frac{1}{2}} d \varepsilon
\end{equation}
and a simple integration furnishes

\begin{equation}\label{e10}
\rho = \rho_q [1-(1-q){\phi\over\sigma^{2}}]^{\frac{5-3q}{2(1-q)}}
\end{equation}
where the density scale is expressed in terms of the complete Beta
function by
$\rho_q=\rho_1{(1-q)^{-\frac{3}{2}}}B(\frac{3}{2},\frac{2-q}{1-q})$.
In the limit $q=1$ one finds $\rho_q = \rho_1$ with the power-law
becoming the exponential, and as expected the standard Maxwellian
result is readily recovered. Substituting (\ref{e10}) into
Poisson's equation, and introducing the pair of dimensionless
quantities defined by $x = r/r_o$ and $\theta =
-\sigma^{-2}\phi(r)$, where $r_o ={\sqrt 4\pi G \rho/\sigma^{2}}$,
we find the Lane-Emden type equation

\begin{equation}\label{e11}
\frac{1}{x^2}\frac{d}{dx}(x^2 \frac{d\theta}{dx})= -[1 +
(1-q){\theta}]^{\frac{5-3q}{2(1-q)}}
\end{equation}
whose importance for stellar dynamics is largely known. First, we
notice that in the limit $q=1$ the above equation reduces to

\begin{equation}\label{e12}
\frac{1}{x^2}\frac{d}{dx}(x^2 \frac{d\theta}{dx})=-e^{\theta}
\end{equation}
which is exactly the Lane-Emden form for a Maxwellian isothermal
sphere (Chandrasekhar 1960, Spitzer 1987, Binney \& Tremaine
1994). The solution of this equation subjected to the boundary
conditions $\theta={d\theta}/{dx}=0$ at $x=0$ has been numerically
computed by Chandrasekhar and coworkers (see Chandrasekhar 1960
and references there in). Some important theorems has been proven
both for the Maxwellian isothermal sphere and the Lane-Emden
equation. Such a study can also be extended to equation
(\ref{e11}) adopting the same boundary conditions.


Now, it should be recalled that the gravitational potential is
defined up to a constant value in an arbitrary level. Therefore,
one may also choose $\phi_o$ in such a way that $\phi \rightarrow
\phi - \phi_o$ subjected to the restriction that both the
potential and the density goes to zero at infinite. With that
choice, the arbitrary constant for $q\neq 1$ is $\phi_o=1/(1-q)$
and from (\ref{e10}) we obtain the following expression

\begin{equation}\label{e13}
\rho = \rho_q {\theta}^{\frac{5-3q}{2(1-q)}}
\end{equation}
whereas the differential equation (\ref{e11}) becomes

\begin{equation}\label{e14}
\frac{1}{x^2}\frac{d}{dx}(x^2 \frac{d\theta}{dx})=-{\theta}
^{\frac{5-3q}{2(1-q)}}
\end{equation}
which is the canonical form of a polytropic Lane-Emden equation of
index

\begin{equation}\label{e15}
n=\frac{5-3q}{2(1-q)}.
\end{equation}


In figure 1 we display the profile $\rho/\rho_q$ for some values
of the $q$ parameter. The labels under each curve correspond to
the value $10q$ and since the density profiles are identical to
the classical polytropic solution we obtain the known restriction
that the total mass diverges above the polytropic index $n=5$
since the Lane-Emden spheres have finite mass only for $n \leq 5$.
In terms of the $q$-parameter this condition is simply translated
as $q \leq 5/7$. On the other hand, from the second law of
thermodynamics we know that $q \geq 0$ (Lima et al 2001).
Therefore, models with finite mass and sizes are obtained if the
nonextensive parameters lies on the interval $0 \leq q < 5/7$.
Such a limit was previously obtained by Plastino and Plastino
(1993) maximizing Tsallis entropy. The unique difference is that
it was referred to as 9/7 because the distribution with cut-off
was written as q-1 instead of 1-q. In fact, substituting $q
\rightarrow 2-q$ we see that $5/7$ goes to $9/7$, with the latter
becoming a lower bound as should be expected. Note that the
physical lower bound constraint ($q \geq 0$) means that $n \geq
5/2$. Therefore, Tsallis distributions whose total energy has a
cut-off ranges only a half of all possible Lane-Emden
distributions with finite mass. However, since the above
Lane-Emden type equations are still valid for $q > 1$, one may see
that $n \geq 0$ means $q \geq 5/3$, and in the limit $q
\rightarrow \infty$ the index $n \rightarrow 3/2$. Hence, we
conclude that Tsallis power law statistics do not describe the
class of polytropic distributions contained on the range $3/2 < n
< 5/2$ since this range requires negative values of $q$. These
results can be seen directly in the plane (n,q) represented in
figure 2.


At light of such results, one may be tempted to conclude that a
subset of Lane-Emden stellar polytropes can be regarded to as
nonextensive collisionless isothermal spheres, thereby giving a
kinetic justification to this class of stellar configurations. It
is argued here, however, that such an identification is misleading
since polytropic models are defined by one of their equation of
state, namely: $p\rho^{-\gamma}=const$, or equivalently,
$T\rho^{1-\gamma} =const$, where $\gamma = 1 + 1/n$ is the
polytropic index. Hence, the temperature in polytropic models
depends on the position, whereas the class of nonextensive spheres
(including the Maxwellian case) has constant temperature which can
formally be defined by the velocity dispersion $\sigma =
\sqrt{k_BT/ m}$.

In figure 3 we display the projected mass density $\Sigma$ for the
same set of the q-parameter illustrated in figure 1. These
profiles illustrates the surface density profile that should be
obtained in an astronomical object described by this class of
Tsallis models.

It is worth notice that Tsallis isothermal spheres have a fraction
of the stars with positive energy and even so they are bounded to
the main structure whose total mass is finite for $q < 5/7$. In
this concern, one may ask what happens if the energy is
constrained by $\varepsilon \leq 0$ as usually adopted for bounded
structures. Physically, one would expect finite structures for a
larger range of the q-parameter. This problem it will be discussed
next section.


\section{Truncated Nonextensive Stellar Models}

Let us now consider a class of nonextensive stellar models which
is obtained by assuming that only objects with $\varepsilon \leq
0$ are present in the distribution. It can be viewed as a simple
generalization of the truncated model proposed by Wooley (1954).
The corresponding density profile as a function of the
gravitational potential assumes the form (Cf. equation (9))

\begin{equation}\label{e16}
\rho= {4\pi{\rho_1 C_{q}}\over {{(2\pi
\sigma)}^{3/2}}}\int_{{\phi\over \sigma}^{2}}^0
[1-(1-q)\varepsilon]^{\frac{1}{1-q}}(\varepsilon -{\phi\over
\sigma^2})^{\frac{1}{2}} d \varepsilon
\end{equation}
where integration limits are defined by equation (\ref{e8}). As
one may check, this integral can be expressed in terms of
hypergeometric functions in the form

\begin{equation}\label{e17}
\rho
=\frac{C_q(\eta(\varphi))^{\frac{1}{1-q}}}{\eta_{*}^{\frac{3}{2}}}
\frac{2}{3} (- \frac{\varphi}{\eta_*})^{\frac{3}{2}}
F(\frac{3}{2},-\frac{1}{1-q},\frac{5}{2}, -
\frac{\varphi}{\eta_*})
\end{equation}
where $\eta(\varphi) = 1 - (1- q)\varphi$,\, $\eta_* =\frac{ 1 -
(1- q)\varphi}{(1- q)}$, and $\varphi = {\phi \over \sigma^{2}}$.
In order to work out a numerical solution it is more convenient to
express this density profile in a form where the integration
limits are fixed as

\begin{equation}\label{e18}
\rho=C_q(-\varphi)^{3/2}\int_0^1[1-(1-q)\varphi\chi]^{\frac{1}{1-q}}(1-\chi)^{\frac{1}
{2}} d \chi
\end{equation}
and one may verify that at the outer halo region, where $|\varphi|
<< 1$, the density profile behaves like

\begin{equation}\label{e19}
\rho \simeq \rho_q
(-\varphi)^{3/2}\;\;\;\;\;\;\;\;\;\;\;\;\;\;\;\;
\;\;\;\;\;\;\;\;\;\;\;\;\;\;\;\;\;\;\ |\varphi|<<1
\end{equation}
corresponding to a polytropic spherical profile having Lane-Emden
index $n=3/2$. In the central region, where we might adopt the
approximation $|\varphi|>>1$, the density profile reduces to the
expression

\begin{equation}\label{e20}
\rho = \rho_q (-\varphi)^\frac{5-3q}{2(1-q)}
\end{equation}
which corresponds to a Lane-Emden index $n=(5-3q)/2(1-q)$.
Therefore, if all the stars has velocity smaller than the escape
velocity, the general density profile in terms of the
gravitational potential can be described as a smooth transition
between two different polytropic spheres, where the case $n=3/2$
behaves like an attractor for the outer halo region. This
behaviour is illustrated in figure 4. As a further consequence all
these truncated models have finite mass extent due to the very
steep density profile dominating the external region.


The density profile of these truncated models can be easily
obtained by a numerical integration of the the Poisson equation
subject to the auxiliary condition given by equation \ref{e17}.
The integration begins at $r=0$ by specifying a value $\varphi_0$
for the gravitational potential at the center of the mass
distribution of the model. Knowing the potential we estimate the
central density predicted by equation \ref{e17}. Due to the
continuity equation the density gradient at the center is null,
and the same happens with the gradient of the gravitational
potential. Once the density and the gradient of the potential are
determined we apply a discretization algorithm to the second order
Poisson equation for a grid of radial points. The algebraic
approximation is used to predict the gravitational potential, and
its gradient, in the next point of the grid. The whole process is
iterated by solving the density in the next point of the grid and
applying recursively the discretization of the Poisson equation.
The overall numerical integration stop at the point where the
gravitational potential is null. At that point the corresponding
density is also null and we reach the surface of the truncated
model.

In figure \ref{fig5} we present the result of applying this
integration algorithm for the cases $q=1$, $0.9$ and $0.8$. In
each panel the continuous line corresponds to the non truncated
model. This model corresponds to the asymptotic case when the
central gravitational potential is sufficiently high to sample the
polytropic structure. In the particular case of $q=1$ we recover
the classical isothermal sphere. On the other extreme we have the
situation where the central gravitational potential tends to zero.
In that case we can see from figure \ref{fig4} that the structure
is described by a $n=3/2$ politrope independently from the $q$
value. For that reason the density profiles for the three models
in figure \ref{fig5} are exactly the same when $\varphi_0
\rightarrow 0$. For intermediate values of the central potential
the models tend to be quite different depending on the exact value
of $q$. In that figure we have chosen in each panel two values of
$\varphi_0$ corresponding to intermediate density profiles.


The density profiles can be integrated along the line of sight so
that we can estimate the projected mass density. Assuming that the
light trace the mass and that the mass to luminosity ratio is
constant we can scale this to the surface brightness. In figure
\ref{fig6} we present this surface brightness as a function of the
$r^{1/4}$ radial coordinate. The symbols are the same as in the
previous figure. Again we can see that large values of $\varphi_0$
tend to be closer to the nontruncated polytropic models as
indicated by the continuous line. When $\varphi_0 \rightarrow 0$
the models are described by the surface mass density obtained from
the $n=3/2$ polytrope.

\section{Final Comments}

Two simple applications of Tsallis' power law kinetic distribution
have been discussed. The proposed kinetic models represent the
mass distribution of collisionless gravitational stellar systems.
The use of the Tsallis distribution introduces an extra parameter
measuring to what extent the velocity distribution departs from
the standard Maxwell-Boltzmann law which is normally used to
represent these objects. It has been proposed as a viable and well
grounded alternative to the standard equilibrium approach in the
presence of long range forces as happens in the astrophysical
context.

As we have seen, the Poisson equation combined with Tsallis
distribution usually reproduces spherically symmetric structures
resembling the classical polytropic spheres. If the natural
cut-off is imposed ($q<1$), the polytropic index $n$ is closely
related to this Tsallis parameter. The basic result is that models
with $q<5/7$ have finite mass while above this limit the mass is
divergent. The radial and projected density profiles were obtained
by solving numerically the Poisson equation for a large range of
the nonextensive parameter. Moreover, if objects with positive
energies are excluded, one can build a larger set of truncated
models with finite mass and sizes. For this class of models the
corresponding profiles were also numerically determined (see
figures 5 and 6).  For both cases, the predicted profiles should
be compared with that ones observed for globular clusters. In
particular, since energy truncated models are presumed to describe
the relaxed state of clusters, it is interesting to investigate
their connection with possible nonextensive extensions of the
spherically symmetric Michie-King models. Finally, we also remark
that the possibility to include elliptical galaxies in this
generalized framework cannot be discarded. These issues will be
discussed in a forthcoming communication.

{\bf Acknowledgements:} The authors are grateful to Raimundo Silva
and Janilo Santos for helpful  discussions. This work was
partially supported by Pronex/FINEP (No. 41.96.0908.00) and CNPq
(62.0053/01-1-PADCT III/Milenio).

\newpage
\begin{figure}[h]
\centerline{\psfig{figure=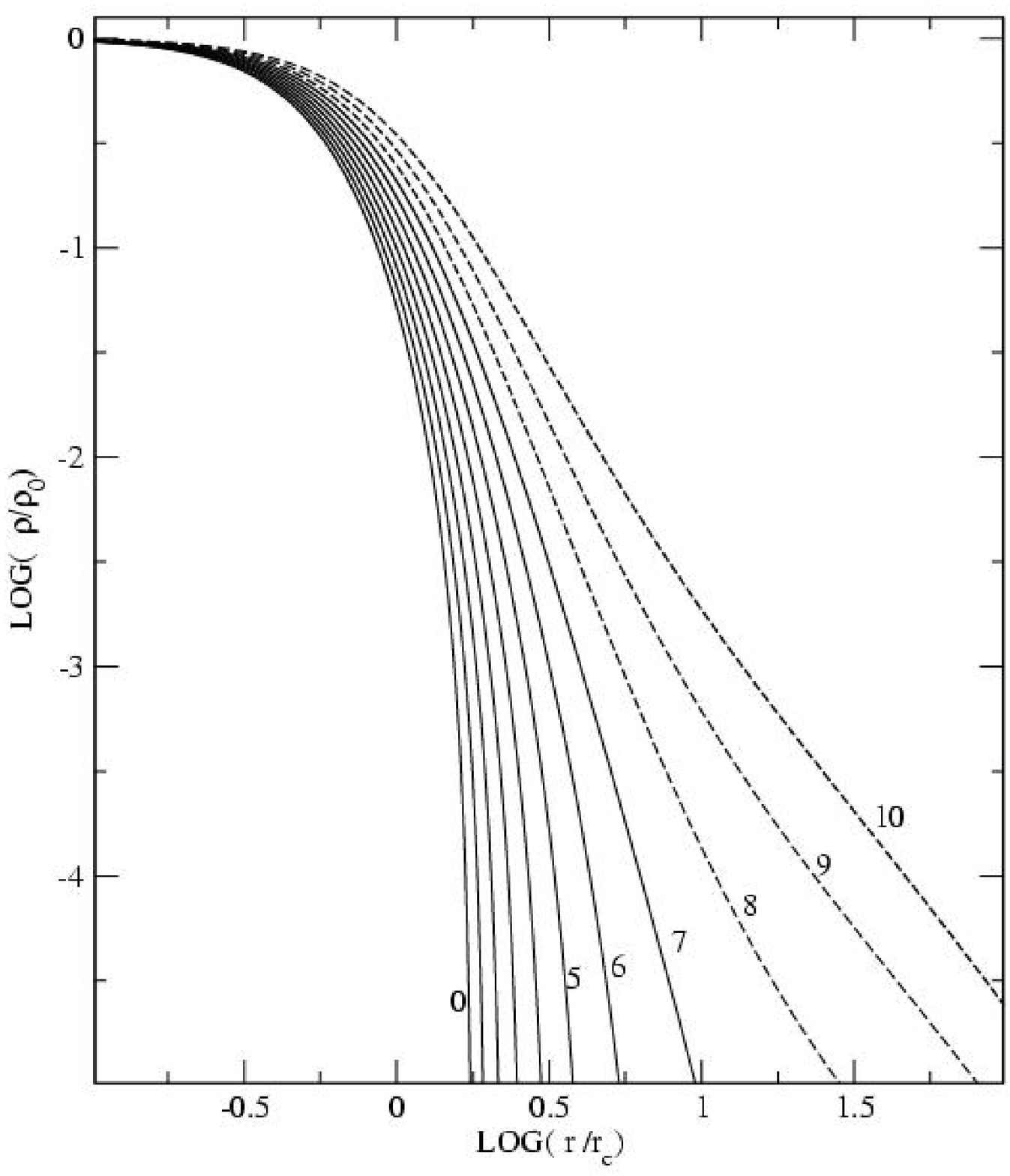,width=8truecm,height=8truecm}
\hskip 0.1in} \caption{Spatial density profile for different
values of $q$. For $q<5/7$ all profiles have finite extent and
finite mass, while for $5/7<q<1$ the models extend to infinite and
the total mass diverges. The limiting value $q=5/7$ corresponds to
$n=5$ polytropic model.} \label{fig1}
\end{figure}

\begin{figure}[h]
\centerline{\psfig{figure=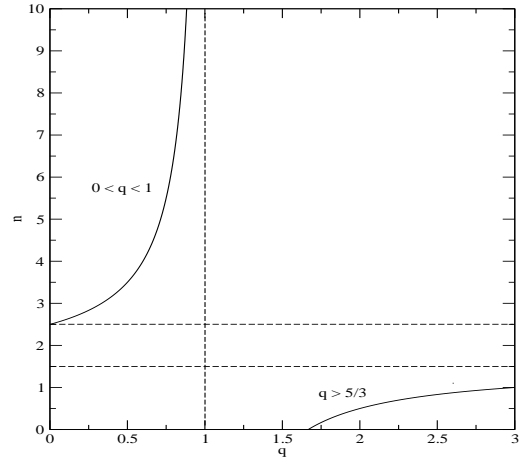,width=8truecm,height=8truecm}
\hskip 0.1in} \caption{The $q$ index of the Tsallis isothermal
description versus the polytropic index $n$. In the left region we
have $0 \leq q <1 $ and the corresponding polytropic models are
restricted to the interval $5/3 \leq n < \infty $. The polytropic
models $0\leq n \leq 3/2$ are described by Tsallis models having
$q>5/3$. Note that polytropic models defined by $3/2 <n<5/2$ do
not have a corresponding Tsallis model.} \label{fig2}
\end{figure}

\begin{figure}[t]
\centerline{\psfig{figure=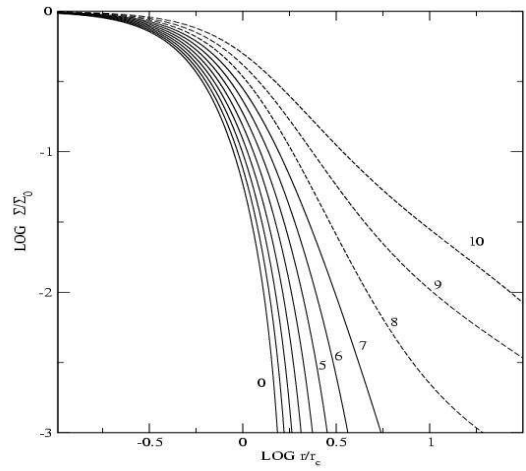,width=8truecm,height=8truecm}
\hskip 0.1in} \caption{Projected mass distribution for different
values of $q$, obtained by numerical integration of the profiles
presented in figure 1.} \label{fig3}
\end{figure}

\begin{figure}[t]
\centerline{\psfig{figure=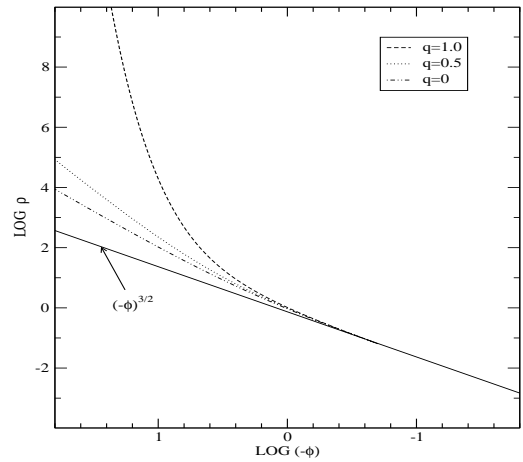,width=8truecm,height=8truecm}
\hskip 0.1in} \caption{A class of truncated models can be obtained
by extracting from the density distribution all objects with
positive energy. In that case the density profile changes and can
be described by a superposition of two limiting polytropic
models.} \label{fig4}
\end{figure}

\begin{figure}[h]
\centerline{\psfig{figure=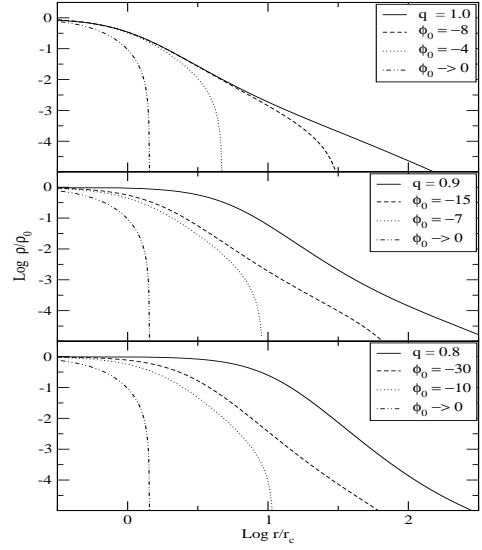,width=8truecm,height=8truecm}
\hskip 0.1in} \caption{Density profiles predicted for the
truncated model corresponding to $q=1$, $0.9$ and $0.8$. The
continuous lines in each panel describes the nontruncated
polytrope corresponding to the limiting case of large $\varphi_0$.
When $\varphi_0 \rightarrow 0$ the models approximate the limiting
$n=3/2$ polytrope independently of the value of $q$.} \label{fig5}
\end{figure}

\begin{figure}[h]
\centerline{\psfig{figure=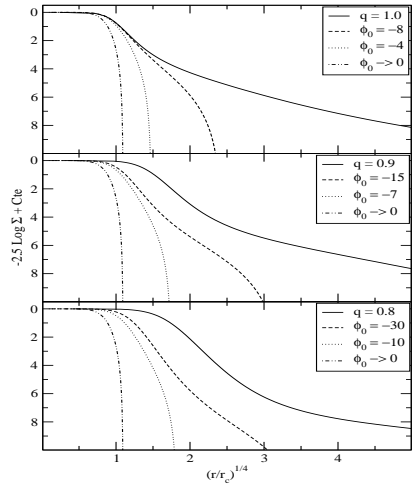,width=7truecm,height=7truecm}
\hskip 0.1in} \caption{Projected mass density for each of the
density profiles of figure \ref{fig5}. } \label{fig6} \end{figure}

\end{document}